\documentclass{appolb}
\usepackage{graphicx}

\begin{document}
\title{Low-energy $\eta$-nuclear scattering parametres
{\it vs.} binding properties in light nuclei
\thanks{Presented at II International Symposium on
Mesic Nuclei, Krak\'{o}w, Poland, September 23, 2013}
}
\author{J. A. Niskanen
\address{Helsinki Institute of Physics, PO Box 64, FIN-00014
University of Helsinki, Finland}
}
\maketitle
\begin{abstract}
The connection of the binding energy and width of possible
bound $\eta$-mesic states is given to the complex scattering
length for $s$-states in the hope that, with knowledge
of the final state interaction this could be useful in searches
of these states. In spite of the strong direct potential
dependence of both observables this connection is seen
to be very model independent even for various nuclei once
the influence of also the effective range is considered.
The importance of this term is pointed also for data analyses.
Although the nucleus considered here is $^{12}$C, extension
to other nuclei is implied in the background work.

\end{abstract}
\PACS{21.85.+d; 24.10.Ht; 13.75.-n}

\section{Introduction}
As discussed in some other talks in this symposium,
low-energy final state scattering parametres have
been the target of intense analyses of $\eta$-meson
production in recent years. Mostly the investigations
aim at the extraction of the complex scattering length
$a$, where the real part is connected to the existence
or nonexistence of a bound state. The definition common
in meson physics
\begin{equation}
q \cot \delta = \frac{1}{a} + \frac{1}{2} r_0 q^2
\label{definition}
\end{equation}
requires a negative real part and
$|a_{\rm R}| > a_{\rm I}$ (more precisely
including also the effective range
$ \mathcal{R} [a^3(a^\ast - r_0^\ast)] > 0$
\cite{he3,c12}).  However, unfortunately
the scattering cross section data alone cannot give just
the sign of this quantity.

Nevertheless, the magnitude of the energy
of a bound state (or a virtual state) can be estimated
from final state interaction (FSI) effects, though
the existence cannot. This was the original idea of
Ref. \cite{he3} studying the relation
in $^3$He {\it provided}
a bound state would exist. This work was recently extended
to heavier nuclei
\cite{c12}, where the existence of bound states
is less controversial and experiments are being performed
(see {\it e.g.} talks by Gal, Machner and Wilkin in these
proceedings).

Particularly suggestive is the possibly rapid change of the
scattering lengths in $^3$He and $^4$He discussed
by Machner at the first symposium of this series
\cite{machner}. There the latter scattering length
appears significantly smaller than the large
$a_{^3{\rm He}\eta}$, which would hint to a change
of the sign (of the real part), since one would
expect more attraction in the heavier nucleus and
thus either even larger scattering length (for the
nonbinding situation) or a binding strength with undetermined
size of the cross section and scattering length.
However, the results seem to depend strongly on whether they
are obtained with or without the effective range term of
the low-energy expansion. This is discussed in subsec.
\ref{efrange}

The aim of this talk is to review and summarize
the numerical relation of the complex binding energy
to the complex scattering length and effective range.
At the end some speculative extrapolation possibility
is shortly played with.

\begin{figure}[tb]
\centerline{%
\includegraphics[width=12.5cm]{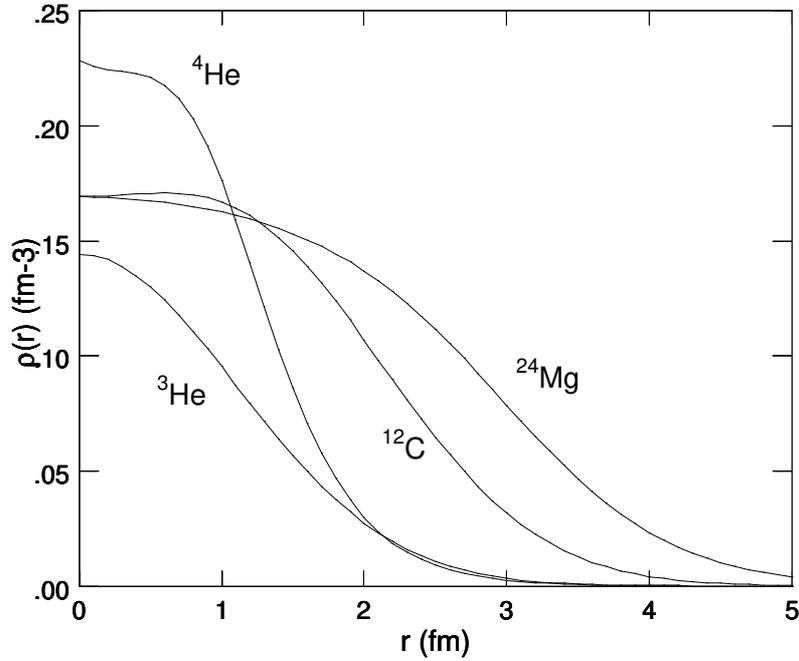}}
\caption{The density profiles of the nuclei used in Refs.
\cite{he3} and \cite{c12}.}
\label{profiles}
\end{figure}

\section{Formalism}
The basic idea is to employ various nuclear charge
distributions $\rho(r)$ in simple optical potentials
\begin{equation}
V_{\rm opt} = -4\pi (V_{\rm R} + i V_{\rm I}) \rho(r)/
(2\mu_{\eta N})\, ,
\label{optical}
\end{equation}
where for each nucleus the strength parameters
($V_{\rm R}, V_{\rm I}$) are
freely varied to get a sufficient coverage to present
the binding energy and width as contours in the
$(a_{\rm R}, a_{\rm I})$ plane.  
The normalization of the profiles shown in Fig.
\ref{profiles} is chosen to the mass number as
$4\pi\,\int_0^\infty \rho(r)\, r^2\, dr = A$, so that
in the simple impulse approximation optical model
the strength
$(V_{\rm R},V_{\rm I})$ would correspond to the
elementary scattering length $a_{\eta N}$
($r$ given in fm). However,
it should be stressed that no claim is made about
any absolute strength of the potential
as in microscopic model works. The main purpose is to get
{\it the numerical connection of binding energies and widths
to the scattering parameters}, so that if the latter can be
extracted from data, then a preliminary estimate could be
obtained for the former. Possibly this would be useful
for planning experiments. In the spirit of the
shape independence of $NN$ forces, one might expect a
more density profile independent connection between these
quantities than in the direct relation to the potential
of either of them.

\begin{figure}[tb]
\centerline{%
\includegraphics[width=12.5cm]{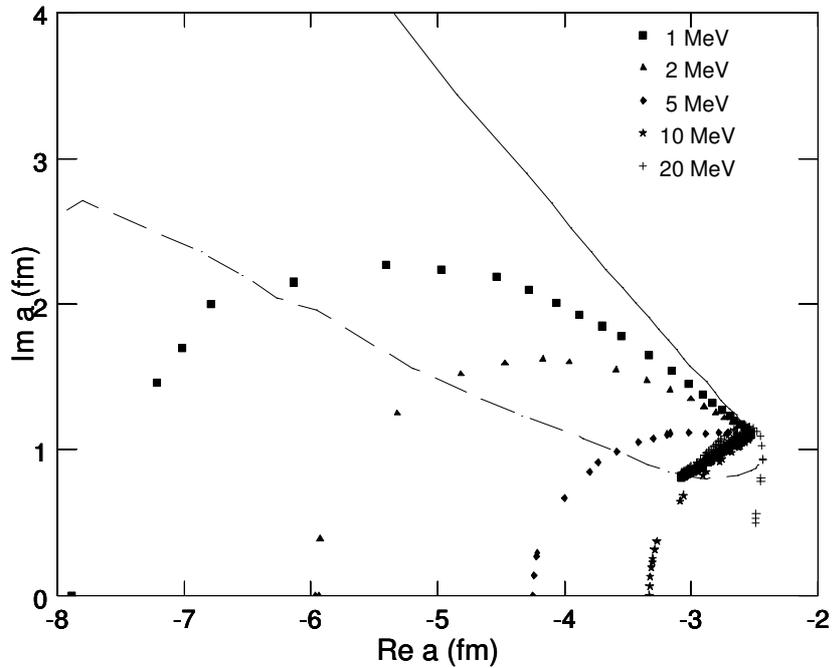}}
\caption{The $s$-wave binding energy
$E_{\rm B} = - E_{\rm R}$ contours for 1, 2, 5, 10 and
20 MeV in the complex $(a_{\rm R},a_{\rm I})$ plane.
The solid line shows the zero energy, {\it i.e.} above it there
is no binding. Below the dashed line $|E_{\rm R}| > |E_{\rm I}|$.}
\label{Ereal}
\end{figure}

If the effective range term is included, then also the
energy relation should include this and the low (complex)
energy relation would be
\begin{equation}
1/a = -\sqrt{-2\mu_{\eta A}E/\hbar^2}
    -  r_0\, \mu_{\eta A}\, E/\hbar^2
\label{relation}
\end{equation}
with $\mu_{\eta A}$ the reduced mass of the system. In
comparison with the actual exact results from solving equations
of motion this relation was found to be much more accurate
than keeping  the first term alone.
For binding energies up to $|E|\approx 10$ MeV and even
beyond the accuracy of eq. \ref{relation}
was a few percent for the real part and
about ten percent
for the imaginary ($E$ and $r_0$ taken from the
calculation). This success can be attributed to the inclusion
of a non-zero effective range $r_0$ and can be considered as
an indication of the necessity of its inclusion also in
data analyses.

\begin{figure}[htb]
\centerline{%
\includegraphics[width=12.5cm]{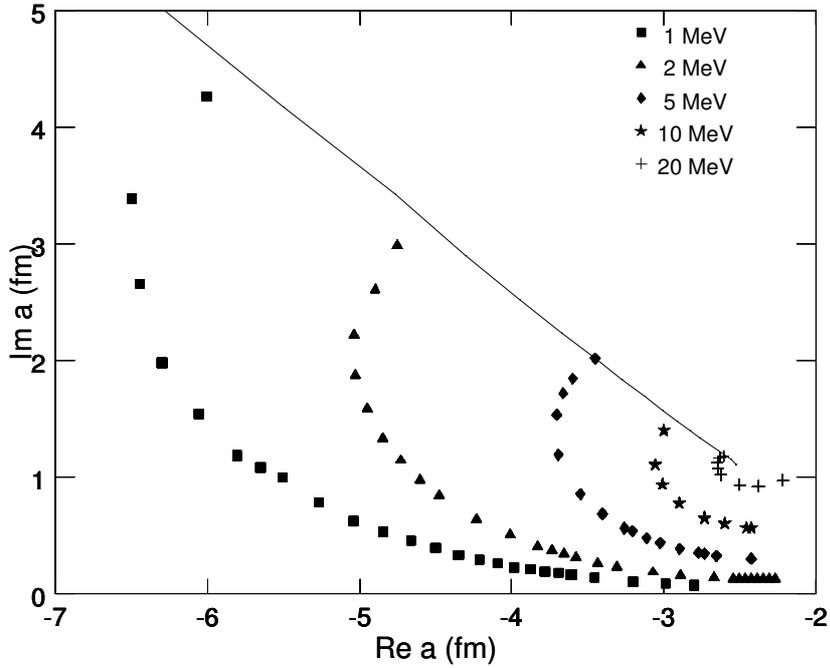}}
\caption{The same as Fig. \ref{Ereal} but for the imaginary part
 of the bound state energies $-E_{\rm I}$, {\it i.e.} half widths.
}
\label{Eimag}
\end{figure}

\section{Numerical results}
\subsection{Case of $^{12}C$}
Going to heavier nuclei from  $^3$He carbon may be a
representative compromise between the experimentally studied
helium and magnesium isotopes. Fig. \ref{Ereal} shows the
binding energy contours for five choices up to 20 MeV.
Also shown are the zero energy limit and the borderline where
$|E_{\rm R}| > |E_{\rm I}|$. The latter is of interest as
a measure of actually distinguishing any peak. To facilitate
the complete connection Fig. \ref{Eimag} shows also the
contours of constant half-widths ($-E_{\rm I}$).

The behaviour for a real potential is quite clear and
understandable, but some peculiar features appear for
strongly absorptive interactions.
Absorption, described by the imaginary part,
acts in the bound system like a repulsion eating
away the wave function in the attractive region: less
wave function $\longrightarrow$ less attraction. So
for increasing absorption the strength of also the
attractive real
part must be increased to stay on the equal-energy contours.
However, as shown in Ref. \cite{c12} the effect of the
imaginary part of the potential saturates in the
scattering length and $a_{\rm I}$ is not even monotonously
increasing. Therefore, the calculated points for the
equidistant mesh of the potential strength become denser
and denser with increasing absorption (and stronger
real part to remain at the same binding energy) and
eventually the contours turn back to left. However,
in this region the states are so wide, anyway, that they
may not be of experimental interest.
Furthermore, it may be worth noting that in the proximity of
zero binding the widths tend to be large, significantly
larger than the binding energy for the same $V_{\rm R}$
without imaginary part at all. Therefore, near threshold
bound states may be hard to distinguish.

\begin{figure}[tb]
\centerline{%
\includegraphics[width=12.5cm]{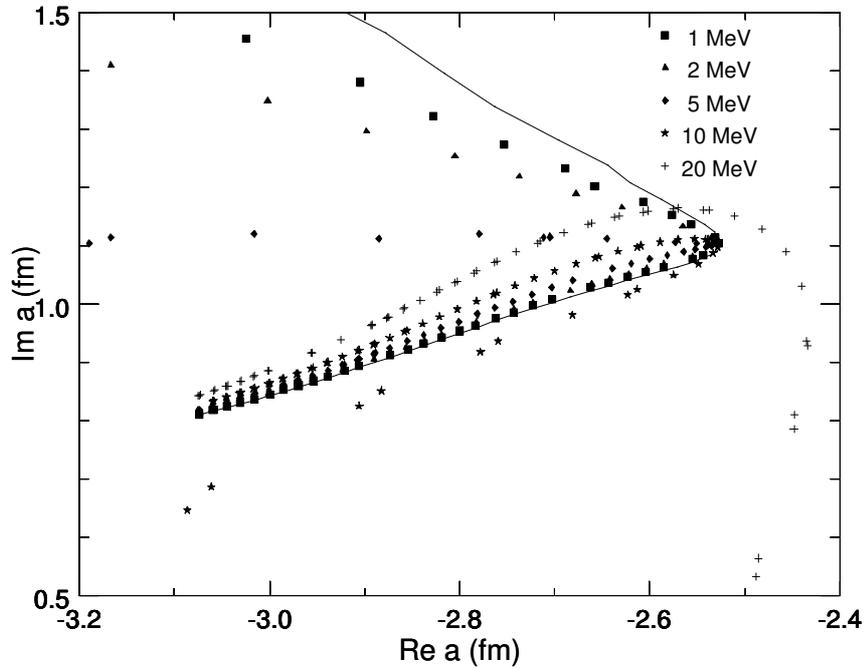}}
\caption{Magnified detail of Fig. \ref{Ereal}
}
\label{mini}
\end{figure}

To get a more detailed picture of that behaviour and
the general trend of increasing binding energies Fig.
\ref{mini} shows a magnified detail of Fig. \ref{Ereal}.
It can be seen that the "weak potential branch" comes
lower and lower in the $a_{\rm I}$ direction
and the opening angle
between the "branches" decreases until they cross
over at about 9 MeV binding. So the 10 MeV contour
below the formal strong potential zero line belongs
actually to the weaker couplings and 20 MeV has gone
even further.

The behaviour seen in Figs. \ref{Ereal} and \ref{Eimag}
is very similar to that of $^3$He \cite{he3} and
also $^4$He and $^{24}$Mg \cite{c12} especially for
large $|a|\; (> 4)$ and "weak potential branch"
dictated mainly by just the binding energy term in
Eq. \ref{relation}. The next section discusses the
potential dependence through the effective range,
which in turn depends naturally on the size of the
nucleus.

\subsection{Significance of the effective range}
\label{efrange}
From relation \ref{relation} it is clear that at least
with "realish" (dominantly real) quantities for
a negative $E$  (consequently also negative $a$)
and a  positive effective range the two terms on the
left hand side tend to cancel making $1/a$ smaller and
$a$ bigger. Due to this cancellation $a$ is quite
sensitive on the inclusion of the second term.
This expectation is actually seen in drastic changes
of the data analyses from
Refs. \cite{Smyrski07,Mersmann07,Budzanowski09b} giving
the scattering lengths shown in Table \ref{scattlen}.
Clearly leaving the effective range out may give
scattering lengths even less than one half of their true
values and consequently grossly overestimated binding
energies. This is confirmed in the calculations and
by Eq. \ref{relation}.

Consequently, it is clear that
consistency in estimating the possible binding energy
from the more advanced scattering analyses needs also the
effective range term. If one takes for example the
scattering lengths for $\eta^4$He above, leaving that
away and making the standard estimate for possible
binding energy, the first one (without $r_0$) would give
about 4 MeV and the second  (with $r_0$) 1 MeV. However,
from Fig. \ref{Ereal} the results
for these two $a$'s would be 10 MeV and
2 MeV, correspondingly, the second being preferable.
(The result for $^4$He, Fig. 8 of Ref. \cite{c12}, is
very nearly the same as $^{12}$C for such weak binding
due to shape independence. Practically only the region
of the "turning point" tends to move left towards larger
$|a_{\rm R}|$ with increasing $A$ and right for decreasing.)

Because of the apparent importance of the effective range
Ref. \cite{c12} gives a parametrization of the complex $r_0$
in terms of the real and imaginary parts of the
scattering length.

\begin{center}
\begin{table}[tb]
\centering \caption{Scattering lengths in fm from data with
and without the effective range term.}
\vspace{3mm}
\label{scattlen}
\begin{tabular}{l|c|c|c|c|c|c}
System  &  $a_{\rm R}$ & $a_{\rm I}$  Ref.\\
\hline
$\eta(^{3}$He) &  $\pm(2.9 \pm 2.7)$ &  $3.2 \pm 1.8 $&
 \cite{Smyrski07} \\
$\eta(^{3}$He) & $\pm (10.7\pm 0.9)$ & $1.5\pm 2.6$ &
\cite{Mersmann07} \\
$\eta(^{4}$He) &  $\pm(3.1 \pm 0.59$ &  $0 \pm 0.5 $&
 \cite{Budzanowski09b} \\
$\eta(^{4}$He) & $\pm (6.2\pm 1.9)$ & $0.01\pm 6.5$ &
\cite{Budzanowski09b} \\
\hline
\end{tabular}
\end{table}
\end{center}

\subsection{Prospect of extrapolation}
\begin{figure}[htb]
\centerline{%
\includegraphics[width=12.5cm]{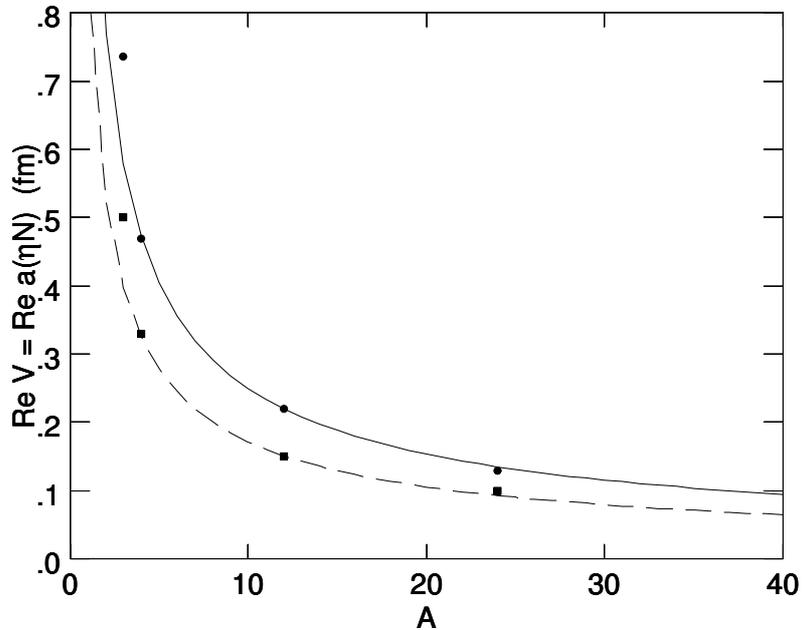}}
\caption{Dependence of the binding limit on the
mass number. Squares and dashed line for real
potentials, dots and solid curve for $V_{\rm I}
= V_{\rm R}/2$.
}
\label{Adep}
\end{figure}

Although the aim of this talk is not to promote
a definite potential model, it may still be tempting
to speculate by interpreting the strength parametres
$V_{\rm R}$ and $V_{\rm I}$ in Eq. (\ref{optical})
to be the real and imaginary parts of the basic
$\eta N$ scattering length $a_{\eta N}$ in a first
order optical potential as done in {\it e.g.}
Ref. \cite{Wilkin}. (Or they might represent
 some more general strengths
perhaps extracted from microscopic nuclear models.)
Then the strength $V_{\rm R}=0.33$ fm, $V_{\rm I}=0$
fm would support a barely bound state in $^4$He,
which would mean that all
standard values of Re $a_{\eta N}$ ranging roughly
between 0.4 fm and 0.7 fm could have the potential
for binding in this nucleus. For this basic strength
the binding energy
in $^{12}$C would be 8 MeV and in $^{24}$Mg 14 MeV.
To produce a barely bound state in the three nuclei
considered above the fitted strength
$V_{\rm R}=0.86\times A^{-0.7}$ fm would be needed
as shown by the squares and the dashed curve in Fig.
\ref{Adep} (also $^3$He is shown).

However, $a_{\eta N}$ is complex and as another
example, assuming more realistically
$V_{\rm I} = 0.5\, V_{\rm R}$
(roughly true for most $a_{\eta N}$in the
literature), $V_{\rm R} = 0.47$
fm would just bind the $\eta ^4$He system and the
corresponding binding energies would be 15 MeV and
22 MeV for $^{12}$C and $^{24}$Mg , respectively.
Their half-widths would be about the same as their
binding energies; in the case of $^4$He the
half-width would be 12 MeV.
In this case the barely binding strength could be
parametrized as $V_{\rm R} = 1.25\times A^{-0.7}$
fm (the circular dots and the solid curve).
It may be of further interest to remind that the
depth of a square well potential producing a zero
binding energy should behave like $R^{-2} \propto
A^{-2/3}$.

\section{Conclusion}
In this work a phenomenological connection between the
low energy scattering length and the complex binding
energy of possible eta-nuclear bound states has been
studied in a simple but probably realistic model.
The hope is that the results could be useful in searches
of these bound states, if more easily accessible
final state data are available to make predictions
where to look for the states. The binding energies are
explicitly presented as contours in the complex $a$
plane for the $^{12}$C (also for $^4$He and $^{24}$Mg
in Ref. \cite{c12}). It is seen that the connection is
very closely the same and systematic for relatively
different density profiles, from which it is easy to
interpolate and even extrapolate to other nuclei.

The calculations suggest that
for even relatively moderate values of the absorptive
potential and of the imaginary parts of the scattering lengths,
the states can be wide especially compared with the
real depths of the states. In view of also many other
theoretical results,
starting from the elementary $\eta N$ scattering and predicting
negative real parts for the scattering length but with rather
large imaginary parts, the observation of such bound states
might be difficult or even impossible. However, in the
minireview \cite{Sibirtsev} of the situation a reanalysis
of the existing data on $\eta^3$He final states made
very small values of the imaginary part appear
plausible, so that also the possible bound states may not
necessarily be as wide as most theoretical works would
indicate.

In our work for $a_{\rm I}$ less than 2 fm with $a_{\rm R}$
larger than, say, 5 fm a bound state should be recognizable.
In the case of more likely smaller scattering lengths
$a_{\rm I} < 1$ fm would be necessary. For the assessment
of possible distinguishable bound states Fig. \ref{Ereal}
indicates the region where the half width is less than
the binding energy. Experimentally in this respect the
result $a_{\rm R} = \pm 6.2 \pm 1.9$ fm and
$a_{\rm I} = 0.001 \pm 6.5$ fm
for $^4$He of Ref. \cite{Budzanowski09b}
is quite interesting and suggestive. If
$|a_{\rm R}(\eta ^4{\rm He}|$ is really
larger than for $\eta ^3$He scattering, it is
difficult to see how this could be true for a more
attractive but still unbinding system.

The relation between $a$ and $E$ is very robust against
potential differences even between different nuclides
over a wide range once also the non-zero effective
range is taken into account. Therefore, due to this shape
independence one may trust the results to be valid
by interpolation also for the $A = 7$ nuclei of recent
experimental interest \cite{li7}.\\

I thank Pawel Moskal for the invitation and kind hospitality
at the Symposium and HIP for a travel grant.


\begin{thebibliography}{99}
\bibitem{he3} A. Sibirtsev, J. Haidenbauer,
J. A. Niskanen, U. Meissner, {\it Phys. Rev. C} {\bf 70}, 047001
(2004).
\bibitem{c12}
J. A. Niskanen, H. Machner, {\it Nucl. Phys. A} {\bf 902},
40 (2013).
\bibitem{machner} H. Machner, {\it Acta Phys. Polonica B}
{\bf 41}, 2221 (2010).
\bibitem{Smyrski07}
J.~Smyrski, \textit{et~al.},  {\it Phys. Lett.} {\bf B 649}
(2007) 258.

\bibitem{Mersmann07}
T.~Mersmann \textit{et~al.},  {\it Phys. Rev. Lett.} {\bf 98}
(2007) 242301.

\bibitem{Budzanowski09b}
A.~Budzanowski \textit{et~al.} (The GEM~Collaboration)
{\it  Nucl. Phys.}
{\bf A 821} (2009) 193.

\bibitem{Wilkin}
    C. Wilkin, {\it Phys. Rev. C} {\bf 47} (1993) R938.

\bibitem{Sibirtsev}
    A. Sibirtsev, J. Haidenbauer, C. Hanhart, and J. A. Niskanen,
    {\it Eur. Phys. Journal A} {\bf 22} (2004) 495.

\bibitem{li7}
A.~Budzanowski \textit{et~al.} (COSY-GEM collaboration), Phys. Rev.
{\bf C 82} (2010) 041001R.


\end{thebibliography}
\end{document}